# Broadband Photovoltaic Detectors based on an Atomically Thin Heterostructure

*Mingsheng Long[1], Erfu Liu[1], Peng Wang[2], Anyuan Gao[1], Wei Luo[1,3], Baigeng Wang[1*], Junwen Zeng[1], Yajun Fu[1], Kang Xu[1], Wei Zhou[1], Yangyang Lv[4], Shuhua Yao[4], Minghui Lu[4], Yanfeng Chen[4], Zhenhua Ni[5], Yumeng You[6], Xueao Zhang[3], Shiqiao Qin[3], Yi Shi[7], Weida Hu[2*], Dingyu Xing[1] and Feng Miao[1*]*

[1] National Laboratory of Solid State Microstructures, School of Physics, Collaborative Innovation Center of Advanced Microstructures, Nanjing University, Nanjing 210093, China.

[2] National Laboratory for Infrared Physics, Shanghai Institute of Technical Physics, Chinese Academy of Sciences, Shanghai 200083, China.

[3] College of Science, National University of Defense Technology, Changsha 410073, China.

[4] National Laboratory of Solid State Microstructures and Department of Materials Science and Engineering, Nanjing University, Nanjing 210093, China.

[5] Department of Physics, Southeast University, Nanjing 211189, China.

[6] Department of Chemistry, Southeast University, Nanjing 211189, China.

[7] School of Electronic Science and Engineering, Nanjing University, Nanjing 210093, China.

**ABSTRACT:** Van der Waals junctions of two-dimensional materials with an atomically sharp interface open up unprecedented opportunities to design and study functional heterostructures. Semiconducting transition metal dichalcogenides have shown tremendous potential for future applications due to their unique electronic properties and strong light-matter interaction. However, many important optoelectronic applications, such as broadband photodetection, are severely hindered by their limited spectral range and reduced light absorption. Here, we present a p-g-n heterostructure formed by sandwiching graphene with a gapless bandstructure and wide absorption spectrum in an atomically thin p-n junction to overcome these major limitations. We have successfully demonstrated a $MoS_2$-graphene-$WSe_2$ heterostructure for broadband photodetection in the visible to short-wavelength infrared range at room temperature that exhibits competitive device performance, including a specific detectivity of up to $10^{11}$ Jones in the near-infrared region. Our results pave the way toward the implementation of atomically thin heterostructures for broadband and sensitive optoelectronic applications.

Current photodetection technologies rely primarily on separate photoactive semiconducting materials with certain bandgaps corresponding to distinct spectral ranges. The realization of broadband light detection is crucial in many important optoelectronic applications, such as sensing, imaging[1] and communication[2,3]. Transition metal dichalcogenides (TMDs) have recently been demonstrated with remarkable optical[4-19] and electronic[20,21] properties, and they are expected to play an important role as an essential unit in future ultra-scaled and integrated optoelectronic circuits.

However, broadband photodetection based on TMD materials remains a central challenge due to limited spectral bandwidth and reduced light absorption similar to conventional group IV and III-V semiconductors. In the past few years, intensive research efforts have been focused on creating van der Waals junctions by stacking selected two-dimensional materials with an atomically sharp interface[22,23]. This approach offers tremendous opportunities for designing functional electronic[16,24] and optoelectronic[15,25-27] units.

In this work, we present a p-g-n heterostructure formed by sandwiching graphene (g) in an atomically thin p-n junction to realize broadband and high-sensitivity photovoltaic detectors. A schematic drawing is shown in Fig. 1a. Graphene with a gapless bandstructure[28-30] provides an effective absorption material over a wide spectral range. The built-in electric field formed at the depletion region of the p-n junction plays a crucial role in efficiently separating photoexcited electrons and holes and enables broadband photodetection with high sensitivity. Compared to previously proposed broadband solutions based on photoconductive graphene hybrid structures[29], such a design could effectively suppress the dark current and overcome the issues of low specific detectivity and large energy consumption. As a demonstration, we fabricated an atomically thin $MoS_2$-graphene-$WSe_2$ heterostructure with a broadband photoresponse in the visible to short-wavelength infrared range at room temperature; this device exhibited a specific detectivity of up to $10^{11}$ Jones (cm $Hz^{1/2}$ $W^{-1}$) in the near-infrared region, which meets the requirements of many important applications.

Fig. 1b shows an optical image and schematic side view of the van der Waals-assembled MoS$_2$-graphene-WSe$_2$ heterostructure (see Methods for the fabrication details). The large-work-function metal Pd was used for doping electrodes to obtain p-doped WSe$_2$, whereas MoS$_2$ flakes remained n-type due to Fermi level pinning[31], as confirmed by their individual transfer curves (see Supplementary Section 3 for details). The current-voltage ($I_{ds}$-$V_{ds}$) characteristic curves of the heterojunction without light illumination were measured, and the results for back gate voltage $V_g$ values from -60 to 60 V are plotted in Fig. 1c. A rectifying behavior was observed. For this typical junction, $I_{ds}$ increased with increasing $V_g$, indicating a p-type nature. The doping type of the junction can be tuned by adjusting the doping level of the MoS$_2$ and WSe$_2$ flakes. Upon exposure to a focused laser beam (488 nm), we observed a photovoltaic effect resulting in non-zero $I_{ds}$ at zero bias, as shown in Fig. 1d. The output $I_{ds}$-$V_{ds}$ curves were measured at $V_g$ = 0 V with the incidence laser power varying from 0.5 to 25 µW. The short-circuit current ($I_{sc}$) was found to increase linearly with incident power intensity, whereas the open-circuit voltage ($V_{oc}$) (~0.23 V) was found to be independent of the laser power, as shown in the inset of Fig. 1d. Considering that the voltage is applied across an atomically thin scale (~1 nm), the internal electric field is estimated to be extremely strong (~2×10$^8$ V m$^{-1}$). Thus, a p-n junction was formed within the vertically stacked MoS$_2$-graphene-WSe$_2$ heterostructure, as expected.

One of the most compelling features of such a p-g-n structure is the potential for realizing broadband photodetection by using graphene's gapless bandstructure. The main benefits of photodetectors include the photoresponsivity $R$ and specific detectivity $D^*$. Here, $D^*$ is defined by $(A\Delta f)^{1/2}R/i_n$, where A is the effective area, $\Delta f$ is the electrical bandwidth and $i_n$ is the noise current. If the noise from the dark current is a major contribution, $D^*$ can be expressed as $R/(2qJ_d)^{1/2}$, where $J_d$ is the dark current density and $q$ is the electron charge. Fig. 2a presents the photoresponse results of a typical device in the visible to mid-wavelength infrared range at room temperature. Here, the laser wavelength was tuned continuously using a monochromator varying from 400 to 2,400 nm with an interval of 100 nm. The device was tested in ambient air with $V_{ds}$ = 1 V and $V_g$ = 0 V. A remarkably strong response was acquired over the entire spectral range of interest, indicating that the p-g-n heterostructure can provide broadband photodetection, as designed. In the visible region, the measured photoresponsivity $R$ was as high as $10^4$ A W$^{-1}$, and the calculated specific detectivity $D^*$ was as high as $10^{15}$ Jones. As the excitation laser wavelength increased, $R$ and $D^*$ decreased sharply, with $R$ at approximately several A W$^{-1}$ and $D^*$ close to $10^{11}$ Jones in the near-infrared region. $D^*$ continued to decrease slowly and became relatively stable at $10^9$ Jones while approaching a wavelength of 2,400 nm.

The wavelength dependence of the photoresponse can be explained by the broadband absorption of the sandwiched graphene and the limited spectral bandwidth of MoS$_2$ and WSe$_2$ used to build the atomically thin p-n junctions. Similar to conventional semiconductors, TMD materials exhibit no absorption[17] when the photon energy is smaller than the bandgap. In the case of monolayer MoS$_2$ and WSe$_2$, the bandgaps $E_{g1}$ and $E_{g2}$ are 1.88[32, 33] and 1.65 eV[34, 35], corresponding to wavelengths of 660 and 750 nm, respectively. Fig. 2b (left) schematically shows the light absorption of the p-g-n heterostructure in the visible range, in which the photon energy is larger than the bandgaps of the TMD materials. All three layered materials (MoS$_2$, graphene and WSe$_2$) are efficient absorption materials that produce abundant photo-generated free carriers, resulting in a considerably higher photoresponse. When the wavelength is near the infrared region and the photon energy is smaller than $E_{g2}$, the interband absorption of both monolayer MoS$_2$ and WSe$_2$ is forbidden. As shown in Fig. 2b (right), the graphene layer becomes the only light absorption material to generate electron-hole pairs, yielding a relatively smaller photoresponse in the infrared range.

We further investigated the incident light power dependence of the photoresponse in both the visible and near-infrared regions, and the major results are plotted in Figs. 3a and 3b, respectively (measured at $V_{ds}$ = 1 V and $V_g$ = 0 V in ambient conditions). Here, we introduce another key figure of merit: the external quantum efficiency (EQE), which is the ratio of the number of photoexcited charge carriers to the number of incident photons and can be expressed as EQE = ($hcI_P$ /$e\lambda P_I$), where $h$ is the Planck constant, $c$ is the speed of light, and $\lambda$ is the wavelength of the incident laser. In the visible range (using a 532 nm laser), as shown in Fig. 3a, both $R$ (left axis) and the EQE (right axis) increase as the laser power $P_I$ decreases and reach values of up to 4,250 A W$^{-1}$ and 1.0×10$^6$ %, respectively, at a power of 0.2 nW. For such weak incident light, $D^*$ is as high as 2.2×10$^{12}$ Jones, which is comparable to commercial InGaAs photodetectors (~10$^{12}$ Jones with the requirement of cooling to 4.2 K for operation)[36,37]. In the near-infrared region (using a 940-nm laser), a similar trend was observed with relevant results shown in Fig. 3b, in which $R$ is as high as 306 mA W$^{-1}$ at a power of 17 nW. These results imply possible practical applications of p-g-n heterostructures in multiple fields, as demonstrated by a customized room-temperature photoelectric imaging system obtained by replacing the CCD unit of a digital camera with our device and a connected pre-amplifier (Supplementary Section 7).

The structure of the ultra-thin vertical p-g-n junction also plays a crucial role in inducing an efficient photo gain mechanism over the entire spectrum. The photoexcited electron-hole pairs in the strong built-in electric field are spontaneously and rapidly separated[38]. The interlayer inelastic tunneling process occurs due to lateral momentum mismatch[39] in randomly stacked interfaces, which is beneficial for reducing interlayer recombination and increasing the carrier lifetime $\tau_{life}$. The gain $G$ can be expressed as $G = \tau_{life}/\tau_{transit}$, where $\tau_{transit}$ is the carrier transit time over the source-to-drain distance $L$. The time $\tau_{transit}$ is related to the drift velocity $V_d$ and device mobility $\mu$ and can be rewritten as $\tau_{transit} = L/V_d = L^2/\mu V_{ds}$. The gain $G$ can be rewritten as $G = \tau_{life}\mu V_{ds}/L^2$. In our p-g-n structure, $\tau_{transit}$ is much shortened because $L$ is ultimately short in an atomically scale. Thus, the combination of the increased $\tau_{life}$ and reduced $\tau_{transit}$ results in a significantly enhanced photoresponse in the designed p-g-n structure.

In spite of this first demonstration of broadband photodetection based on atomically thin p-g-n junctions, further improvements in device performance can be reasonably expected upon optimization of the device parameters and further engineering of the material properties. For example, the device photoresponse could be enhanced by tuning both $V_{ds}$ and $V_g$ (see Supplementary Section 4 for details). The light absorption of graphene is highly sensitive to the Fermi level, which determines certain allowed interband excitations. Thus, improved device performance could be obtained by engineering the Fermi level to be located at the Dirac point.

To shed light on the photoresponse mechanism of the p-g-n photodetector, we conducted photocurrent mapping measurements by scanning the laser spot over a typical device. Fig. 4a and 4b present the photocurrent mapping at $V_{ds}$ = 0 V and 1 V, respectively, when $V_g$ = 0 V. An optical micrograph of the device is shown in Fig. 4c, which indicates the positions of the electrodes and different layered materials in the photocurrent images (highlighted by different colored lines). The photocurrent mapping measurements were performed under an 830 nm laser at a power of ~20.5 μW under ambient conditions. The photocurrent mappings reveal that the strongest photoresponse occurs within the overlapped region (highlighted by red dashed lines) of $MoS_2$, graphene and $WSe_2$, rather than the electrode regions. This suggests that the photoresponse in our devices originates from the designed atomically thin p-g-n junction, rather than the Schottky barriers formed at the metal contacts[40] or other device regions (see additional data sets in Supplementary Section 5).

In summary, we present a novel p-g-n heterostructure to enable broadband photovoltaic detection. Compared to previously proposed broadband solutions based on photoconductive structures, this design could effectively overcome their limitations of low specific detectivity and large energy consumption. We successfully fabricated an atomically thin $MoS_2$-graphene-$WSe_2$ heterostructure and demonstrated broadband photoresponse in the visible to short-wavelength infrared range at room temperature along with competitive device performance. Our results underscore the great potential of van der Waals junctions based on two-dimensional materials for future important optoelectronic applications.

**Methods**

We used a standard mechanical exfoliation method to isolate mono/few-layer $MoS_2$, $WSe_2$ and graphene films. The thickness of the flakes was first measured using a Bruker Multimode 8 atomic force microscope (AFM) and further confirmed by Raman and photoluminescence measurements (Supplementary Section 2). The vertical atomically thin heterostructures were obtained using a polymer-free van der Waals assembly technique followed by a transfer process to a Si wafer covered by a 300-nm-thick $SiO_2$ layer (Supplementary Section 1). A conventional electron-beam lithography process (FEI F50 with Raith pattern generation system) followed by standard electron-beam evaporation of metal electrodes (typically 30 nm Pd/ 30 nm Au) was used to fabricate the vertical p-g-n heterostructure devices.

Electrical transport measurements were performed using a Keithley 2636A dual channel digital source meter. Photoresponse measurements were performed under ambient conditions using a Lake Shore probe station. The wavelength-dependent photoresponse in Fig. 2a was measured using a supercontinuum full spectrum white light laser source (400-2400 nm) combined with a monochromator. The spot size was ~50 μm.

**ASSOCIATED CONTENT**

**Supporting information**

Experimental details about Van der Waals heterostructure fabrication; material characterization (Raman, AFM and PL spectra); individual $MoS_2$ and $WSe_2$ transfer curves; additional photoresponsivity measurements as function of wavelength, bias and back gate voltage; time-resolution photovoltaic response; additional infrared photocurrent mapping; room-temperature photoelectric imaging. This material is available free of charge via the Internet at http://pubs.acs.org.

**AUTHOR INFORMATION**

**Corresponding Author**
E-mail: miao@nju.edu.cn (M. F.); wdhu@mail.sitp.ac.cn (W. H.); bgwang@nju.edu.cn (B. W.).

**Author contribution**

M. L., F. M. and W. H. conceived the project and designed the experiments. M. L., E. L., P. W., A. G., W. L. and J. Z. performed device fabrication and characterization. M. L., W. H., F. M. and B. W. performed data analysis and interpretation. Y. L. and S. Y. performed the $WSe_2$ single crystal growth. F. M., W. H., M. L. and B. W. co-wrote the paper, and all authors contributed to the discussion and preparation of the manuscript.

**Notes**

The authors declare no competing financial interests.

**ACKNOWLEDGMENTS**

This work was supported in part by the National Key Basic Research Program of China (2015CB921600, 2013CBA01603, 2013CB632700), the National Natural Science Foundation of China (11374142, 11322441, 61574076), the Natural Science Foundation of Jiangsu Province (BK20130544, BK20140017, BK20150055), the Fund of the Shanghai Science and Technology Foundation (14JC1406400), the Specialized Research Fund for the Doctoral Program of Higher Education (20130091120040), and Fundamental Research Funds for the Central Universities and the Collaborative Innovation Center of Advanced Microstructures.

**FIGURES**

Figure 1. Broadband photovoltaic detector based on atomically thin p-g-n structure. a, Schematic drawing of the p-g-n heterostructure for photodetection. b, Top: optical image of a fabricated device based on van der Waals-assembled $MoS_2$-graphene-$WSe_2$ heterostructure. Scale bar, 5 μm. Bottom: schematic side view of the heterostructure. c, $I_{ds}$-$V_{ds}$ characteristic curves without illumination, measured at various $V_g$ (from -60 to 60 V, as denoted by different colors). d, Output $I_{ds}$-$V_{ds}$ curves measured with exposure to a focused 488 nm laser beam (with power varying from 0.5 to 25 μW) at $V_g$ = 0 V. Inset: extracted short-circuit current ($I_{sc}$) and open-circuit voltage ($V_{oc}$) versus incident power intensity.

Figure 2. Broadband photoresponse. a, Photoresponsivity $R$ (left) and specific detectivity $D^*$ (right) of a typical device for wavelengths ranging from 400 to 2,400 nm. The device was tested in ambient air at $V_{ds}$ = 1 V and $V_g$ = 0 V. b, Top: spectrum from ultraviolet to mid-infrared with the bandgap of monolayer $MoS_2$ ($E_{g1}$) and $WSe_2$ ($E_{g2}$) indicated. Bottom left: Schematic band diagrams and light absorption of the p-g-n heterostructure in the ultraviolet and visible range, in which the photon energy is larger than the bandgaps of the TMD materials ($E_{g1}$ and $E_{g2}$). Both the two TMDs and graphene respond to produce photo-generated free carriers. Bottom right: the case in which the wavelength is approaching the infrared region and the photon energy becomes smaller than $E_{g2}$. Only graphene responds to produce photo-generated free carriers.

Figure 3. Illumination-power-dependent photoresponse. Measured photoresponsivity $R$ (left axis) and external quantum efficiency EQE (right axis) of a typical device versus illumination power $P_I$ in the visible region (a, using a 532 nm laser) and in the near-infrared region (b, using a 940 nm laser). Both $R$ and EQE increase as $P_I$ decreases. The measurements were performed in ambient conditions with $V_{ds}$ = 1 V and $V_g$ = 0 V.

Figure 4. Photocurrent mapping of the p-g-n photodetector. a, b, Photocurrent mapping results for a typical device at $V_{ds}$ = 0 V (a) and $V_{ds}$ = 1 V (b) with $V_g$ = 0 V. Measurements were performed under an 830 nm laser at a power of ~20.5 μW in ambient conditions.

The photocurrent mappings show the strongest photoresponse within the overlapped p-g-n region (highlighted by gray dashed lines). c, Optical microscopy image of the device with the measurements performed. The $MoS_2$, graphene and $WSe_2$ are highlighted by yellow, light gray and green dashed lines, respectively. The scale bar is 5 μm.

**TOC Figure**

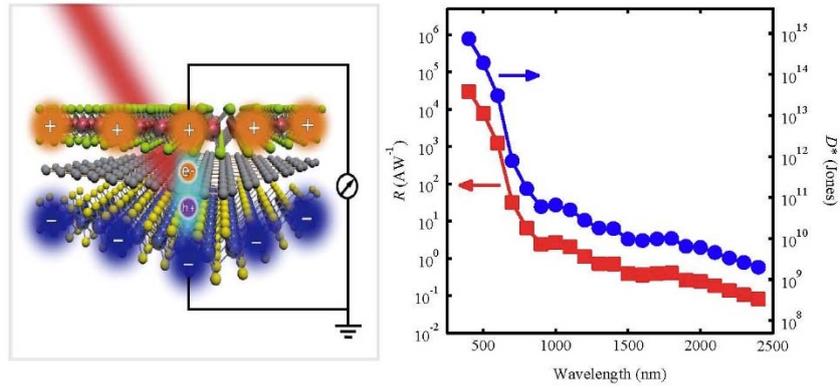

a 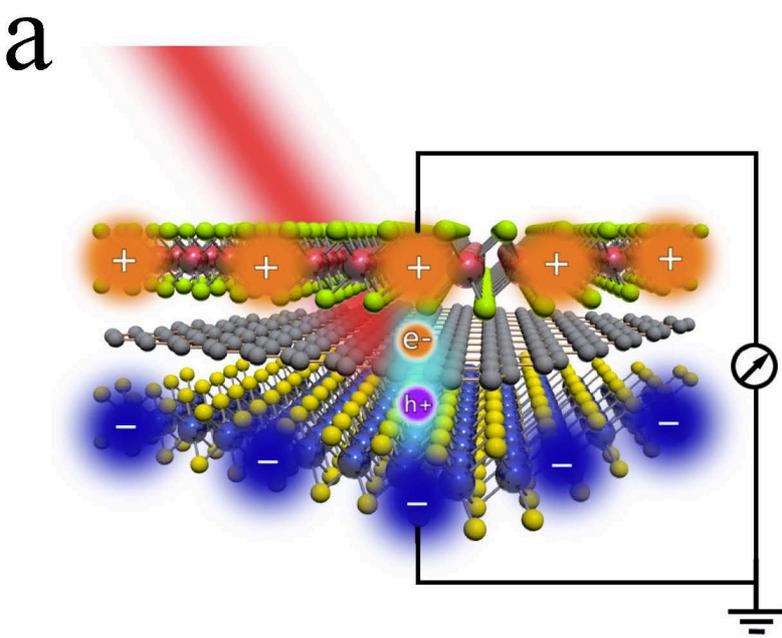

b 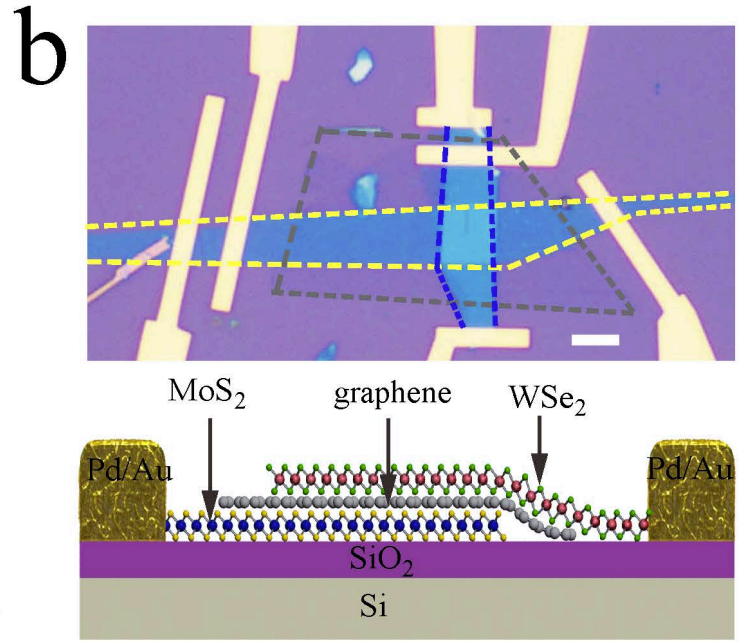

c 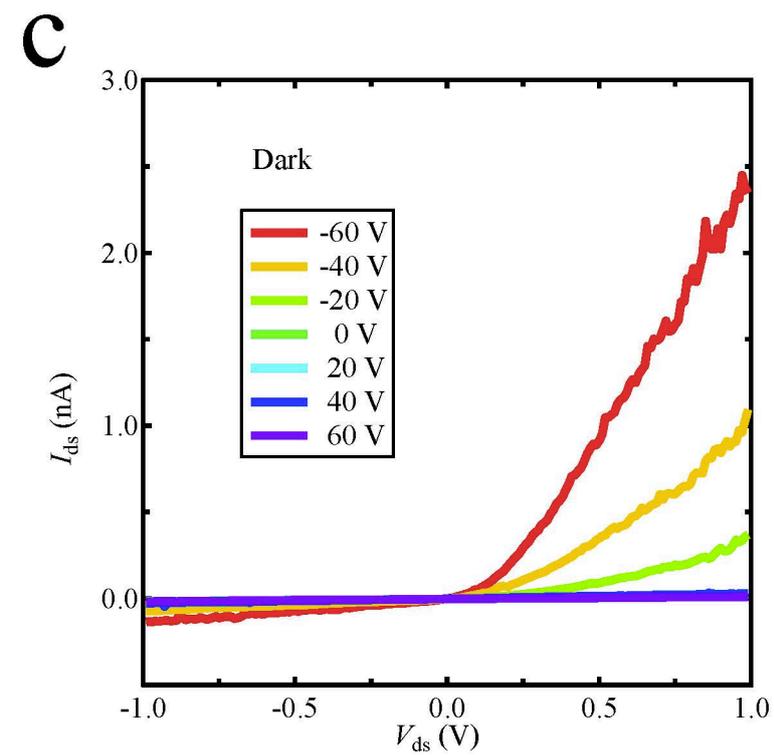

d 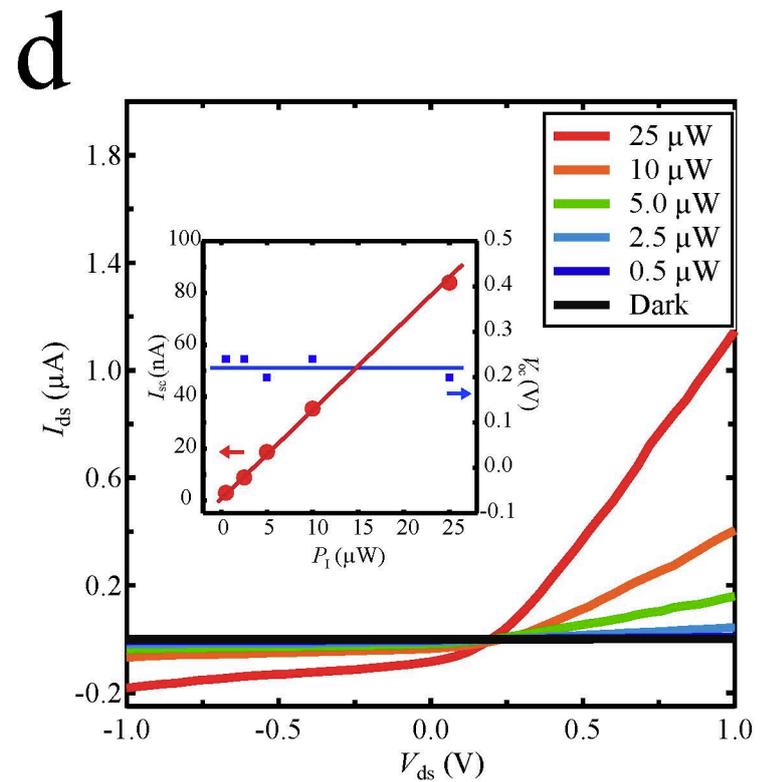

Figure 2 (Long *et. al.*)

a
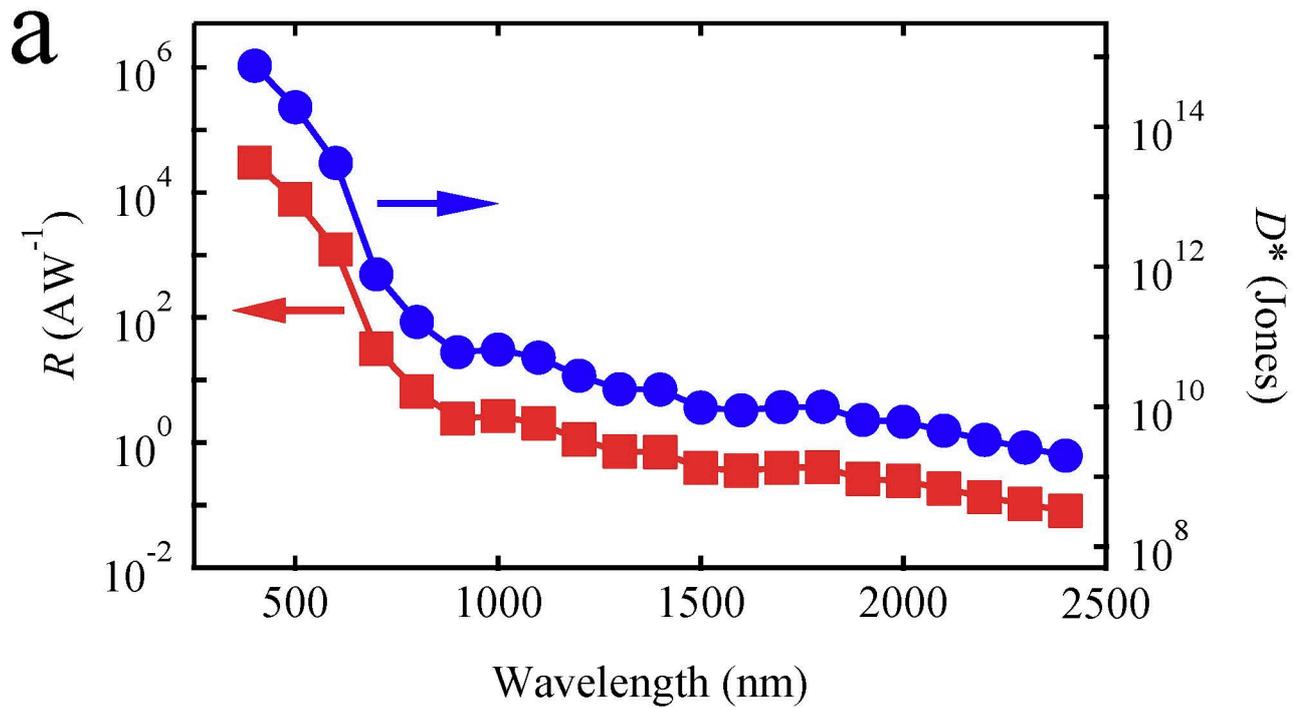

b
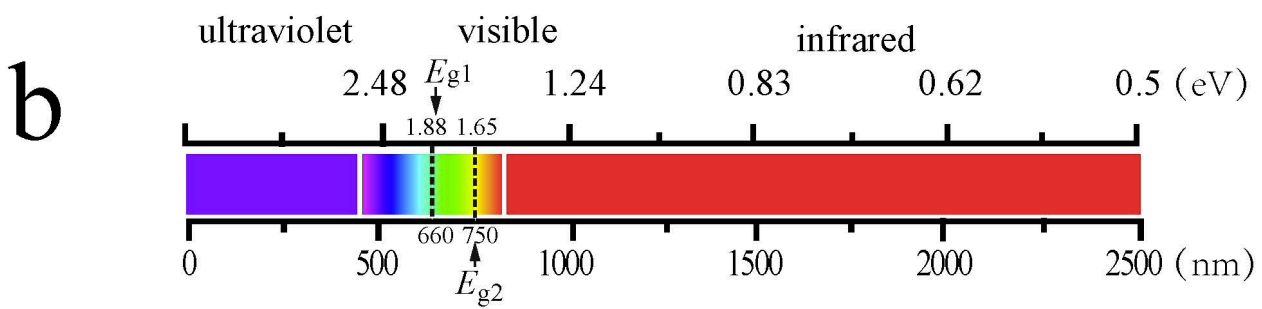

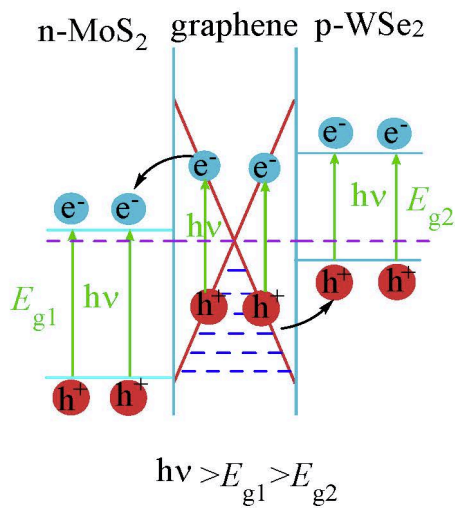 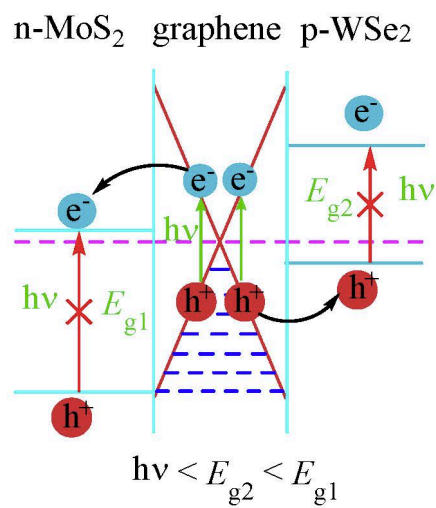

a
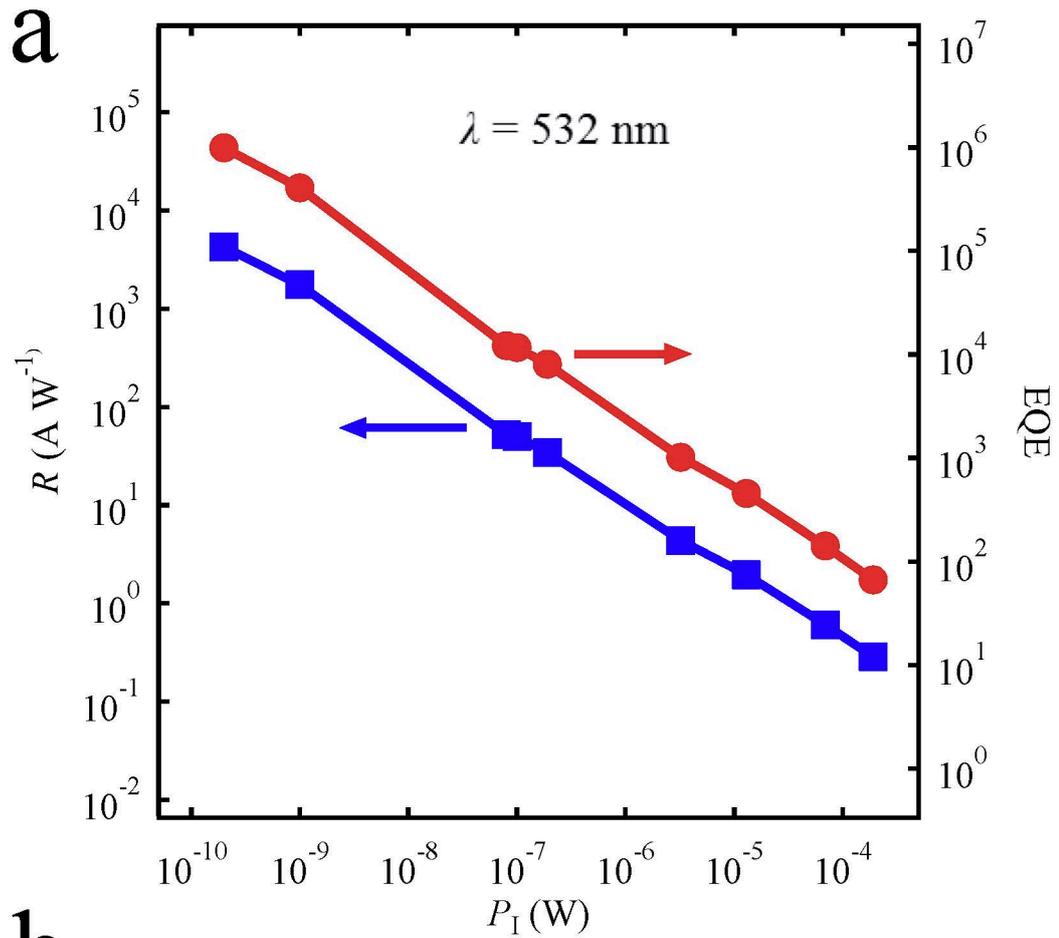

b
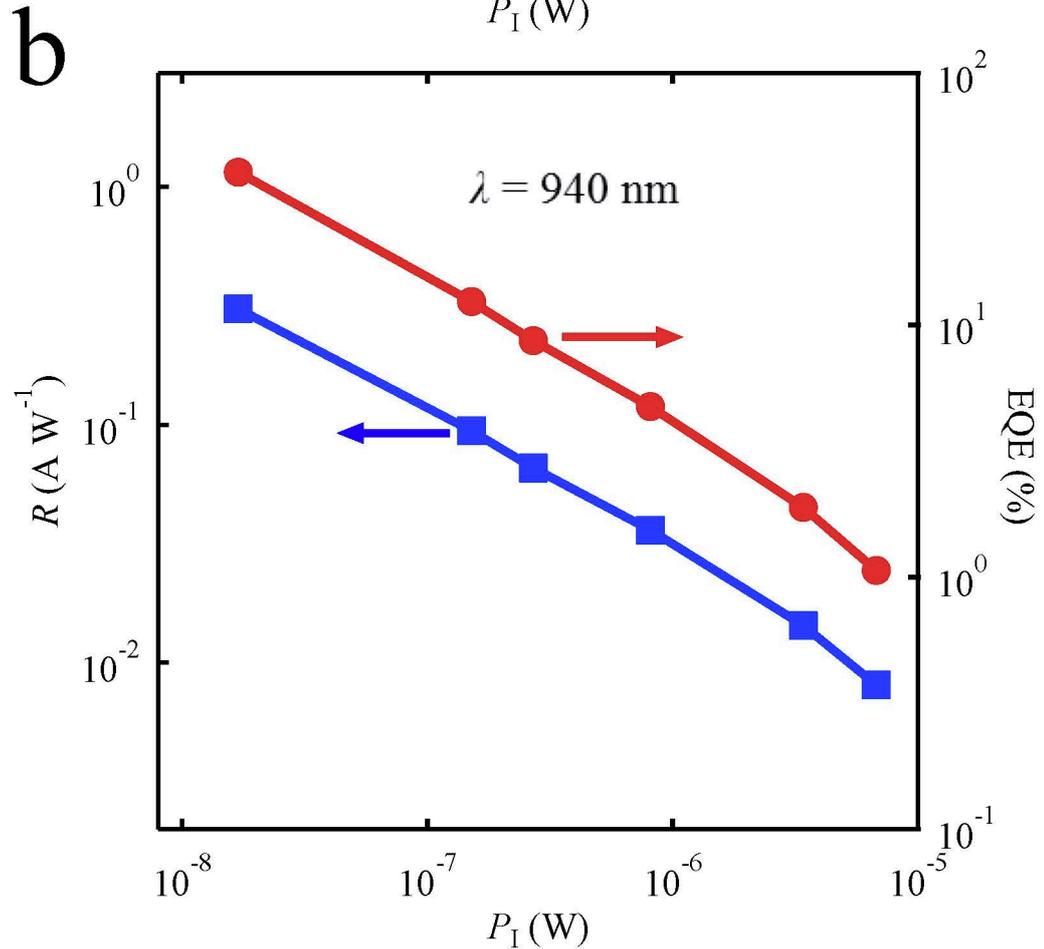

Figure 4 (Long *et. al.*)

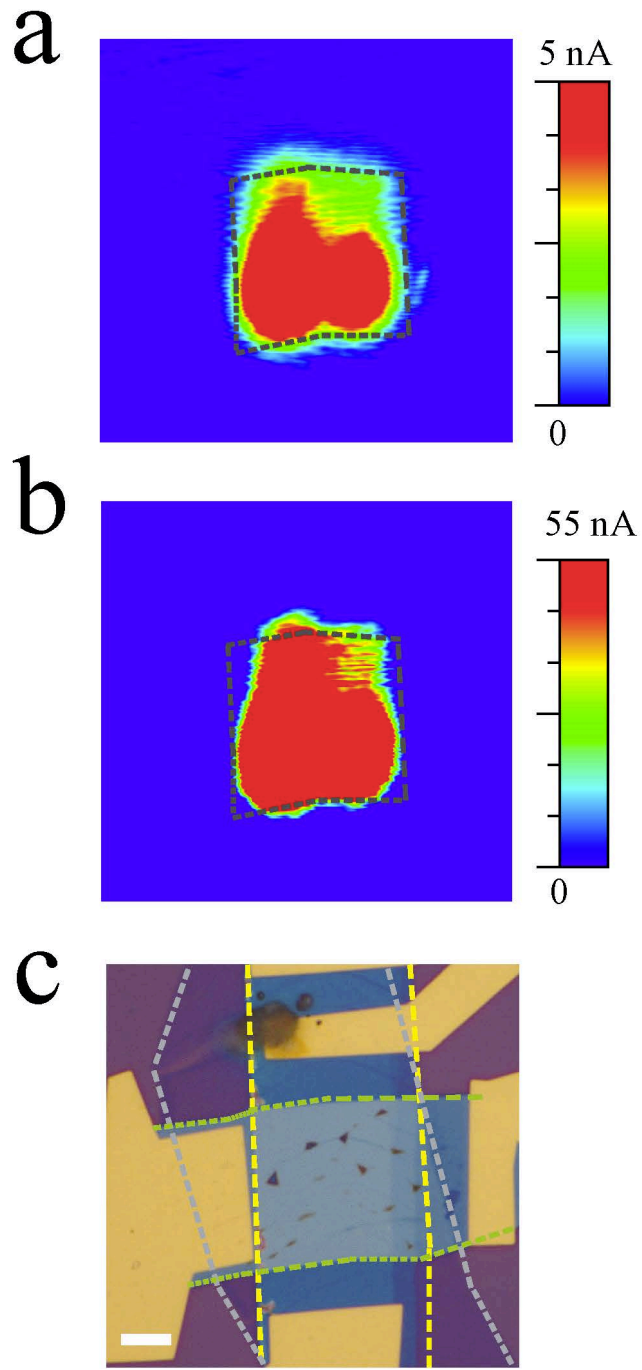

# Supplementary Information For

# Broadband Photovoltaic Detectors based on an Atomically Thin Heterostructure

Long *et al.*

1. Van der Waals heterostructure fabrication
2. Material characterization: AFM, Raman and PL spectra
3. Additional data from electrical characterization
4. Additional photoresponse measurement results
5. Additional data from photocurrent mapping in infrared range
6. Room-temperature photoelectric imaging

1. **Van der Waals heterostructure fabrication**

The p-g-n devices were fabricated by a van der Waals adhesion technique to assemble layered materials without exposing the interfaces to polymers or solvents[1,2]. The $MoS_2$, $WSe_2$ and graphene flakes were mechanically exfoliated from bulk crystals of $MoS_2$ (SPI Supplies), $WSe_2$ and Kish graphite (Covalent Ceramics), respectively, onto a silicon wafer covered by a 285 nm-thick $SiO_2$ layer[3]. Then, a thin layer of water-soluble polyvinyl acetate (PVA) was placed onto a layer of transparent elastomeric stamp poly-dimethyl siloxane (PDMS) to pick up the target flakes. A micromanipulator was used to affix the stamp to certain locations. We started with a blank PDMS/PVA layer to pick up a $WSe_2$ flake from the $SiO_2$ substrate and used the $WSe_2$ flake to pick up a graphene flake via van der Waals adhesion. We then used the $WSe_2$/graphene heterostructure to pick up the bottom $MoS_2$ flake to form the p-g-n structure. The release process (from the PDMS layer) was performed by heating it to 70°C and softening the PVA layer. The PVA was removed in deionized water, and the residual polymers were removed in acetone solution.

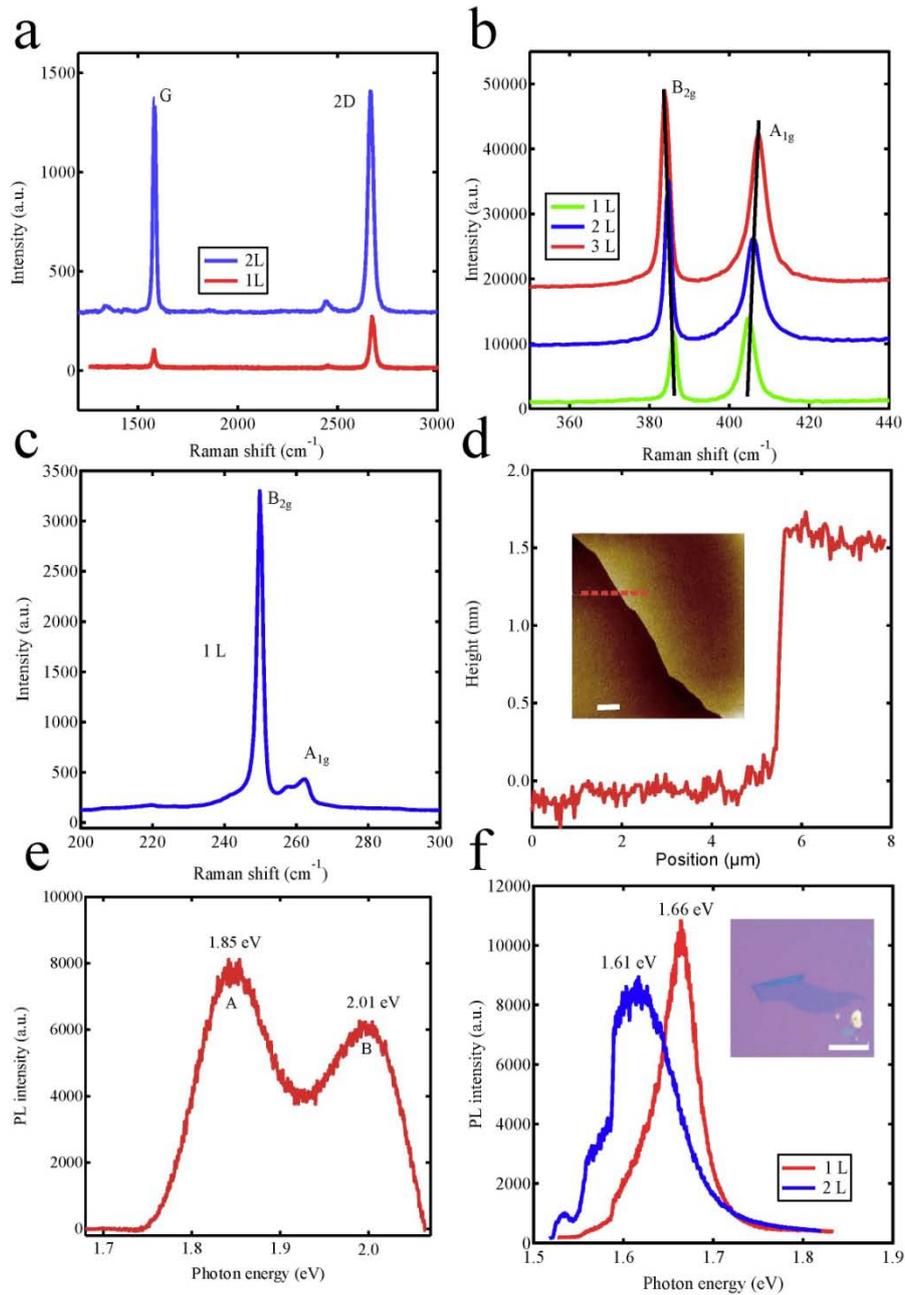

**Figure S1 | Raman spectra, AFM images and PL spectra of the materials. a**, Raman spectra of mono- and bi-layer graphene flakes. The ratio of the 2D to G peak ($I_{2D}/I_G$) ~1 and the 2D band full width at half-maximum ($W_{2D}$) ~30.0 cm$^{-1}$ indicate bi-layer graphene (blue). The $I_{2D}/I_G$ and $W_{2D}$ values of ~3.2 and 26.8 cm$^{-1}$, respectively, indicate monolayer graphene (red). **b**, Raman spectra of mono-, bi- and tri-layer MoS$_2$ flakes. The separation between in-plane $E^1_{2g}$ and out-of-plane $A_{1g}$ vibration modes was found to be 18.5, 21.3 and 23.4 cm$^{-1}$ for mono-, bi- and tri-layer, respectively[6,7]. **c**, Raman spectrum of mono-layer WSe$_2$. **d**, AFM image of a bi-layer MoS$_2$ flake, the

height of which was measured to be approximately 1.5 nm. Scale bar, 2 μm. **e,** PL spectrum of a bi-layer $MoS_2$ flake. The two PL peaks at 1.85 and 2.0 eV correspond to A and B direct excitonic emission, respectively[11,12]. **f,** PL spectra of mono- and bi-layer $WSe_2$. Inset: optical micrograph of a mono-layer $WSe_2$ flake with a scale bar of 20 μm.

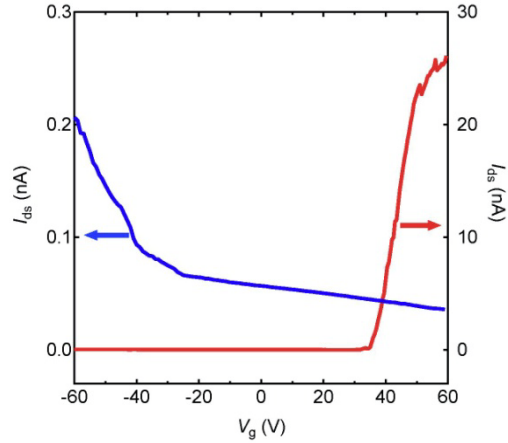

**Figure S2 | Additional data from electrical characterization.** Typical transfer curves of two individual $MoS_2$ (red) and $WSe_2$ (blue) flakes, indicating n-doping and p-doping, respectively. The measurements were performed in the dark at ambient conditions with $V_{ds}$ = 1 V.

2. **Material characterization: AFM, Raman and PL spectra**

Fig. S1 presents the characterization results of Raman spectra, the AFM image and PL spectra for thin graphene, $MoS_2$ and $WSe_2$ flakes with different numbers of layers. The AFM measurements were performed using a Bruker Multimode 8 system. Raman and PL measurements were performed using a Witec alpha 300R confocal Raman system. The laser was focused through a 50× objective lens (NA = 0.7) with a spot size of ~500 nm. The excitation light source was a 532 nm laser (2.33 eV) with a power below 1 mW to avoid sample damage. The Si peak at 520 $cm^{-1}$ was used for calibration of the measurements.

## 3. Additional data from electrical characterization

Fig. S2 shows the transfer curves of the individual $MoS_2$ and $WSe_2$ flakes that were used to assemble the $MoS_2$-graphene-$WSe_2$ heterostructure studied in Figs. 1c and 1d. The $MoS_2$ flake exhibited typical n-type FET characteristics, whereas the $WSe_2$ flake exhibited typical p-type FET characteristics.

## 4. Additional photoresponse measurement results

Fig. S3 presents additional photoresponse measurement results of the photocurrent $I_p$ in the visible to near-infrared range at $V_{ds} = 2$ V and $V_g = 0$ V in ambient conditions. Fig. S4a presents the temporal photoresponse at $V_{ds} = 0$ V and $V_g = 0$ V by varying the laser (532 nm) power intensity. The inset of Fig. S4a presents the photoswitching rate of the detector and displays a fast rising time of ~30 ms. For the telecommunications wavelength of 1.55 μm and mid-infrared wavelength of 2 μm, the time-resolved photovoltaic response at $V_{ds} = 0$ V and $V_g = 0$ V in ambient conditions are plotted in Fig. S4b.

We further studied the bias and back gate dependence of the photoresponse of our p-g-n photodetectors using a 532 nm laser. In Fig. S5a, we extracted $R$ (left axis) and $D^*$ (right axis) for several different $V_{ds}$ values when $V_g = 0$ V; $R$ increased linearly in the forward direction and slightly sub-linearly in the reverse direction, which can be attributed to the asymmetric built-in electric field and interface barriers. In both directions, $D^*$ increased and saturated quickly due to a considerable increase in the dark current when $V_{ds}$ was increased. When $V_g$ was varied from -60 to 60 V ($V_{ds} = 1$ V), $R$ and $D^*$ increased and were as large as ~2.6 A W$^{-1}$ and 2.2×10$^9$ Jones, respectively, as shown in Fig. S5b.

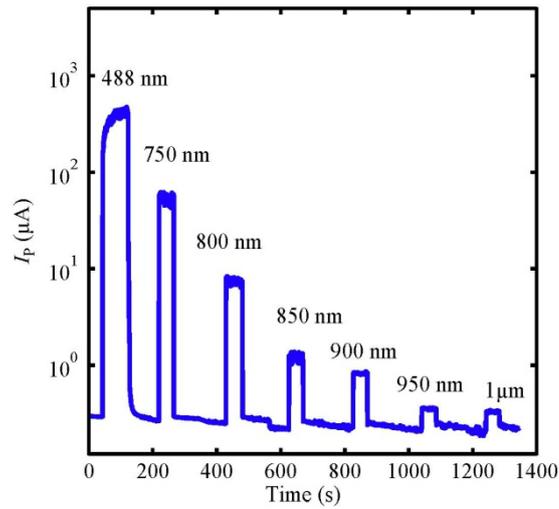

**Figure S3 | Laser-wavelength-dependent photoresponse.** The device was tested at $V_{ds} = 2$ V and $V_g = 0$ V in ambient conditions. The power density was 0.51 W cm$^{-2}$. The spot size of the laser beam was ~50 μm.

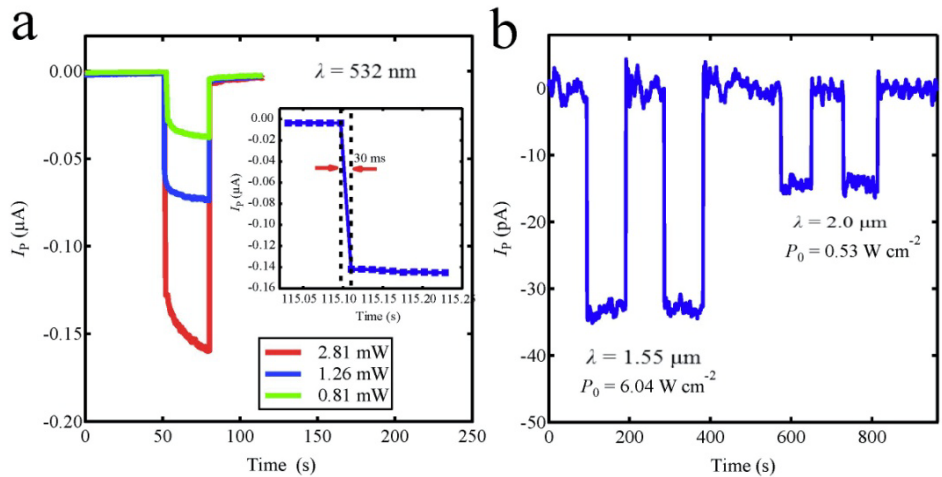

**Figure S4 | Temporal photoresponse. a**, Time-resolution photovoltaic response at different laser (532 nm) powers at $V_{ds} = 0$ V and $V_g = 0$ V under ambient conditions. Inset: Photoswitching rate of the p-g-n photodetector, which exhibits a fast rising time of ~30 ms. The laser power was 2.81 mW. **b**, Temporal response at the telecommunications wavelength of 1.55 μm and mid-infrared wavelength of 2 μm. The device was tested at $V_{ds} = 0$ V and $V_g = 0$ V under ambient conditions. The laser power densities were 6.04 and 0.53 W cm$^{-2}$ for 1.55 and 2.0 μm lasers, respectively.

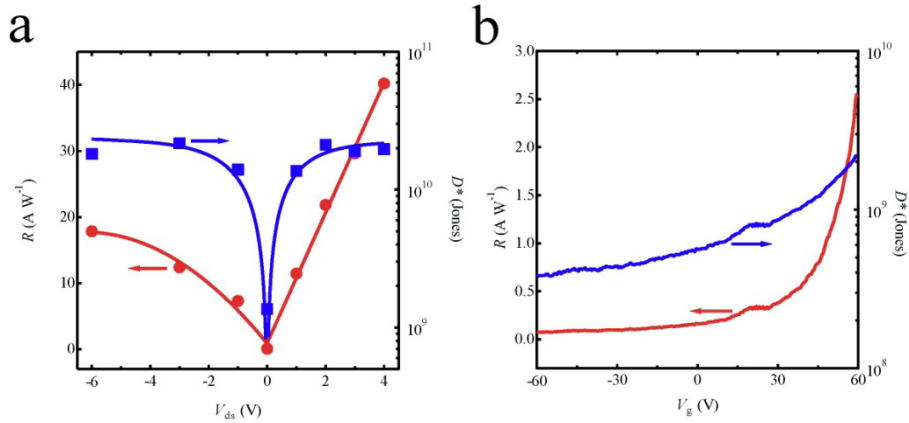

**Figure S5 | Bias and back gate dependence of the photoresponse.** Photoresponsivity $R$ (left axis) and specific detectivity $D^*$ (right axis) as a function of bias voltage (**a**) and back gate voltage (**b**). The incident laser wavelength was 532 nm with a spot size of ~700 nm and excitation power of 1.6 μW.

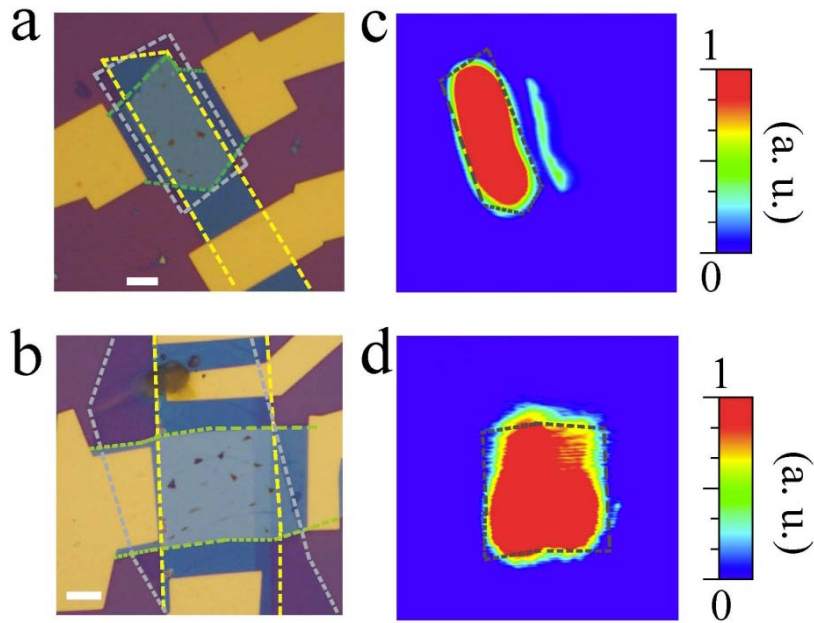

**Figure S6 | Infrared photocurrent mapping. a, b** Optical micrographs of two devices with $MoS_2$, graphene and $WSe_2$ highlighted by the yellow, light grey and green dashed lines, respectively. The scale bar is 5 μm. **c,** The photocurrent mapping result of the device shown in (**a**) corresponds to an 830 nm laser at $V_{ds} = 1$ V and $V_g = 0$ V under ambient conditions. **d,** Photocurrent mapping result of the device shown in (**b**) corresponds to a 940 nm laser at $V_{ds} = 1$ V and $V_g = 0$ V under ambient conditions.

## 5. Additional data from photocurrent mapping in the infrared range

Fig. S6 presents two additional data sets of photocurrent mapping of p-g-n photodetectors in the infrared range. Optical micrographs of two devices are shown in Figs. S6a and S6b, in which different flakes are highlighted by different colored lines. Fig. S6c shows the photocurrent mapping result of the device shown in Fig. S6a under an 830 nm laser (with power ~20.5 μW) at $V_{ds}$ = 1 V and $V_g$ = 0 V under ambient conditions. Fig. S6d shows the photocurrent mapping result of the device shown in Fig. S6a under a 940 nm laser (with power ~19.1 μW) at $V_{ds}$ = 1 V and $V_g$ = 0 V under ambient conditions.

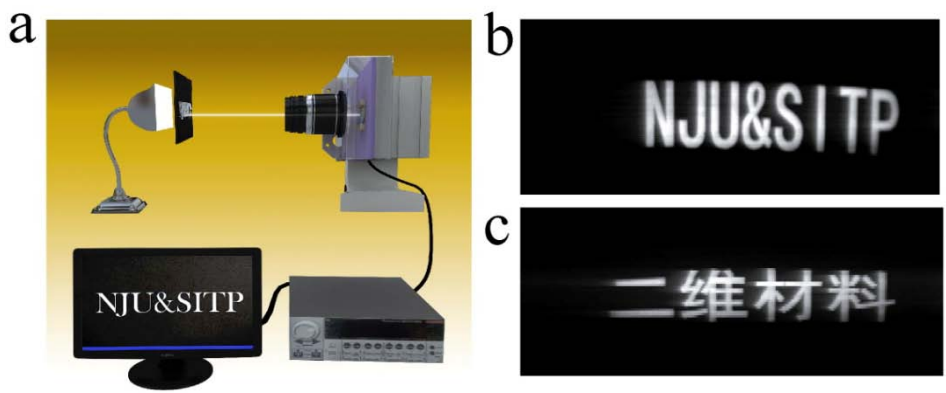

**Figure S7 | Room-temperature photoelectric imaging. a**, Photoelectric imaging system. **b, c,** Obtained images of two printed objects ("NJU & SITP" and "two-dimensional material" (in Chinese)) with good contrast and spatial resolution.

## 6. Room-temperature photoelectric imaging

The realization of broadband and sensitive photodetection of p-g-n heterostructures suggests possible applications in many important fields. Here, as a simple demonstration, we constructed a p-g-n-structure-based imaging system by replacing the CCD unit of a digital camera with our device, which was connected to a pre-amplifier. Fig. S7a schematically shows this integrated imaging system, in which the modified camera was placed on a piezo-electrically controlled platform and the imaging target was a printed object placed in front of a fluorescent light source. Thus, the spatially resolved photocurrent can be recorded as a function of 2D coordinates. Using this imaging system, two images of printed objects ("NJU & SITP" and "two-dimensional material" (in Chinese)) were obtained with good contrast and spatial resolution, as shown in Figs. S7b and S7c, respectively.